\documentclass[twocolumn,prc,superscriptaddress,showpacs,preprintnumbers,amsmath,amssymb]{revtex4}
\usepackage{graphicx}
\usepackage{dcolumn}
\usepackage{bm}
\usepackage{multirow}
\usepackage{hhline}
\usepackage{bbm}

\usepackage{CJK}
\usepackage{color}

\begin{document}

\title{Effect of initial fluctuations on the collective flow in intermediate-energy heavy ion collisions}
\author{J. Wang}
\affiliation{Shanghai Institute of Applied Physics, Chinese
Academy of Sciences, Shanghai 201800, China} \affiliation{University of Chinese Academy of Sciences, Beijing 100049, China}
\author{Y. G. Ma\footnote{Corresponding author. E-mail address: ygma@sinap.ac.cn }}
\affiliation{Shanghai Institute of Applied Physics, Chinese
Academy of Sciences, Shanghai 201800, China}
\affiliation{Shanghai Tech University, Shanghai 200031, China}
\author{G. Q. Zhang}
\affiliation{Shanghai Institute of Applied Physics, Chinese
Academy of Sciences, Shanghai 201800, China}
\author{W. Q. Shen}
\affiliation{Shanghai Institute of Applied Physics, Chinese
Academy of Sciences, Shanghai 201800, China}
\affiliation{Shanghai Tech University, Shanghai 200031, China}
\date{\today}

\begin{abstract}
A systemical analysis of the initial fluctuation effect on the
collective flows for Au+Au at 1$A$ GeV has been presented in the
framework of Isospin-dependent Quantum Molecular Dynamics model (IQMD),
 and a special focus on the initial fluctuation effect on the squeeze-out
 is emphasized. The flows calculated by the participant plane reconstructed
 by the initial geometry in coordinate space are compared with those
 calculated by both the ideal reaction plane and event plane methods. It is found that
 initial fluctuation weakens squeeze-out effect, and some
  discrepancies between the flows extracted by the above different plane methods appear
  which indicate
 that the flows are affected by the evolution of dynamics. In addition, we found
 that the squeeze-out flow is also proportional to initial eccentricity.  Our calculations
 also qualitatively give the similar trend for the excitation function of the
 elliptic flow of the FOPI experimental data. Finally we address the nucleon
 number scaling of the flows for light particles. Even though initial
 fluctuation decreases the ratio of $v_4/v_2^2$ as well as $v_3/(v_1v_2$)
 a lot, all fragments to mass number 4 keep the same curve and shows
 independent of transverse momentum.

\end{abstract}
\maketitle

\section{INTRODUCTION}

Recent relativistic heavy-ion collision (HIC) studies have recognized the
importance of the initial fluctuations on the various order of flows
\cite{Socolowski2004,Andrade2006,
Andrade2008,Takahashi2009,Zhu2005,Petersen2011,Alver2008,Alver2010,XuJ,MaG,Han2011,MaL,ZhangWN}.
For instance,  the origin of
the triangular flow $v_3$ and higher harmonics is fluctuation in initial conditions
 \cite{Han2011,Schenke2011,Takahashi2009}. In particular, $v_3$ vanishes,
 if the system starts with a smooth almond-shaped initial state \cite{Han2011}.
It also shows the ratio of elliptic flow to eccentricity
($v_{2}$/$\varepsilon_{2})$ is sensitive to the initial fluctuation.
By now, studies on the initial fluctuation effects are only limited in relativistic
heavy-ion collision. At these energies, the initial fluctuation can be followed by hydrodynamic
expansion and result in the long-range azimuthal correlations (or the anisotropy flows).
Its energy dependence is thought as one of important observables to study various aspects of the QCD
phase diagram  in the beam energy scan (BES)
program at RHIC \cite{STAR-BES,Tian,NST-Ko,NST-Liu}.
In intermediate energy HIC, no systematic study for the initial fluctuation effects on
the collective flow has been presented except a brief  report appeared by the same authors
of the present work \cite{NST-wj}. Also, the relationship between the eccentricity and
squeeze-out (negative elliptic flow) is not yet reported to our knowledge. In contrast with relativistic energy HIC,
the time scale of collision dynamics is larger (from tens of fm/c to hundreds of fm/c)
in intermediate energy HIC, which may allow the other factors to develop and smear the
long-range azimuthal correlations originating from the initial fluctuation. It is then
worth addressing how the initial fluctuation affects on the collective flow in
intermediate energy domain.

In this paper,  the collective flows  are calculated in a framework of a nuclear transport
model, namely Isospin Quantum Molecular Dynamics (IQMD), which allows the generation of
events with event-by-event fluctuating initial conditions. To explore the effects of the
initial fluctuation, we study the flows as functions of centrality, transverse component
of four velocity and rapidity. Flow results with respect to different reaction plane
determinations are compared with the experimental data. We also try to reproduce
experimental excitation function of elliptic flow with our simulations. In addition,
 the relationship of elliptic flow versus eccentricity
is discussed in different centralities. The EOS dependence of the collective flow is also presented. At the end,
we will focus on phenomenology of mass-number scaling behavior of different harmonic
flows and check the initial fluctuation effect on scaling behavior.

The paper is organized in the following way. A brief description of the IQMD
model is introduced in Sec. II. The initial fluctuation is described in
Sec. III. The methodology of the flow calculations is presented in Sec. IV. The results and
discussion are presented in Sec. V. Finally, summary is given in Sec. VI.

\section{BRIEF DESCRIPTION OF IQMD}

The Quantum Molecular Dynamics (QMD) approach is an n-body theory to simulate
heavy ion reaction at intermediate energies. It contains several major
parts: the initialization of the target and the projectile nucleons, the
spread of nucleons in the effective potential, the collisions between the
nucleons, and the Pauli blocking effect \cite{Aichelin1991}.  There are different versions
of QMD which are best used in different energy regions \cite{Kruse1985,Hartnack1998,Marayama,He,UrQMD}. Of which,
the IQMD model is based upon the QMD model and takes the isospin effects into
account in various aspects:
different density distribution for neutron and proton, the asymmetry
potential term of the mean field, experimental cross-section for
nucleon-nucleon and Pauli-blocking for neutron and proton respectively \cite{Kruse1985,Hartnack1998}.

In IQMD each nucleon is represented by a Gaussian wave packet with a width
parameter (here L = 8.66 fm$^{2})$ centered around the mean position
$\vec{{r}}_{i}(t)$ and mean momentum $\vec{{p}}_{i}(t)$ \cite{Zhang2011,Tao-NST}:

\begin{equation}
\label{eq1}
  \phi_i(\vec{r},t) = \frac{1}{{(2\pi L)}^{3/4}}
exp[-\frac{{(\vec{r}- \vec{r_i}(t))}^2}{4L}] exp[-\frac{i\vec{r}
\cdot \vec{p_i}(t)}{\hbar}].
\end{equation}

The nucleons interact by means of nuclear mean field and nucleon-nucleon
collision. The nuclear mean field can be written by:
\begin{equation}
\label{eq2}
\begin{array}{l}
 U(\rho ,\tau_{z} )=\alpha (\frac{\rho }{\rho_{0} })+\beta
(\frac{\rho }{\rho_{0} })^\gamma+\delta \cdot \ln^{2}(\varepsilon \cdot
(\Delta p)^{2}+1)\cdot \\
\\
\hspace{4.7em}
(\frac{\rho }{\rho_{0} })+\frac{1}{2}(1-\tau_{z} )V_{c}
\frac{(\rho_{n}
-\rho_{p} )}{\rho_{0} }\tau_{z} +U^{Yuk}, \\
 \end{array}\quad
\end{equation}
where $\rho_{0}$ is the normal nuclear matter density (here $\rho_{0}=$
0.17 fm$^{-3}$), $\rho_{n}$ and $\rho_{p}$ are the total, neutron and
proton densities separately, $\tau_{z}$ is $zth$ component of the isospin
degree of freedom, which equals 1 or -1 for neutrons or protons,
respectively. The coefficients $\alpha$, $\beta$ and $\gamma$ are the Skyrme parameters,
which connect closely with the EOS of bulk nuclear matter. Two of them are
fixed by the constraints that the total energy is minimum at the saturation
density $\rho = \rho_{0}$ with a value of $E/A$ = -16 MeV which corresponds to
the volume energy in the Bethe-Weizs\"{a}cker mass formula and the free
particle without binding energy. The third parameter is fixed by the nuclear
compressibility, which is defined by:
\begin{equation}
\label{eq3}
\kappa =9\rho^{2}\frac{\partial^{2}}{\partial \rho^{2}}(\frac{E}{A}).
\end{equation}

Two different equations of state are commonly used: a hard equation of
state (H) with a compressibility of $\kappa$ = 380 MeV and a soft equation
of state (S) with a compressibility of $\kappa$ = 200 MeV \cite{Gregoire1987,Gossiaux1995,Aichelin1987}. $\delta $
and $\varepsilon$ are the optional coefficients for the momentum dependent potential,
which are taken from the measured energy dependence of the proton-nucleus
optical potential \cite{Hartnack1998}. HM and SM mean the hard equation of state and soft equation
of state with momentum dependent potential. C$_{sym}$ is the symmetry energy strength due to
the difference of neutron and proton. $V_{c}$ is the Coulomb potential and
$U^{Yuk}$ is Yukawa potential.

For the collisions, IQMD uses the experimental cross-section which contains
the isospin effects and nuclear medium effect \cite{Hartnack1998}. The Pauli blocking is
also considered after the collisions, to consider fermion property of nucleons.

In IQMD initialization, the centroid of the Gaussian in a nucleus is
randomly distributed in a phase space sphere (r $\le$ R and p $\le$ p$_{F})$
with $R$ = 1.12A$^{1/3}$ fm corresponding to a ground state density of
$\rho_{0}$ = 0.17 fm$^{-3}$ \cite{Voloshin1996}. $P_{F}$ is the Fermi momentum, which
depends on the ground state density. For $\rho_{0}$ = 0.17 fm$^{-3}$ it has
a value of about $P_{F}\approx$ 268 MeV/$c$. In this sense,  the
distribution of the density is nonuniform in IQMD, which leads to the initial
fluctuation. This makes the study of the initial fluctuation effects on the
collective flow in IQMD possible.

\section{Anisotropic flow and reaction plane}

During heavy-ion collisions, the particle azimuthal distribution with
respect to the reaction plane may not be isotropic which is accustomed to be expanded
in a Fourier series \cite{Voloshin2008}:
\begin{equation}
\label{eq4}
E\frac{d^{3}N}{d^{3}p}=\frac{1}{2\pi }\frac{d^{2}N}{p_{T} dp_{T}
dy}(1+\sum\limits_{n=1}^\infty {2v_{n} \cos (n(\varphi -\psi^{RP} ))} ),
\end{equation}
where the $v_{n} = \langle \cos [n(\varphi_{i} -\psi^{RP} )] \rangle$ coefficients are normally
referred as $n$-th collective flow or anisotropic flow \cite{Manly2006} and $\psi
^{RP}$ is the reaction plane angle. Ideally, the reaction plane is defined by the
vector of the impact parameter and the beam direction. It is found that the
magnitude of $v_{n}$ is strongly correlated not only with the initial spatial
eccentricity $\varepsilon_{n}$ but also with the determination of reaction plane.

\subsection{Initial fluctuation and participant plane method}

The \textit{initial fluctuation}, or the fluctuation of the initial collision geometry,
originates from the quantum fluctuation in wave function of projectile and target \cite{Alver2010}.
It affects both the orientation of the reaction plane
and the value of eccentricity. At the moment of the maximum compression
in the HIC, the overlap area is formed, which fluctuates from event to event.
At this moment, the participant plane angle can be defined as:
\begin{equation}
\label{eq5}
\psi^{PP}_{n} =\frac{1}{n}[\arctan \frac{\langle r^{2}\sin (n\phi )\rangle}{\langle r^{2}\cos
(n\phi ) \rangle}+\pi ],
\end{equation}
here $r$ and $\phi$ are the polar coordinate position of each nucleon and the
average $\langle ...\rangle$ is density weighted in the initial state.
The $n$-th collective flow $v_{n}$ with respect to participant plane is
defined as:

\begin{equation}
\label{eq6}
v_{n} =\langle \cos [n(\phi -\psi^{PP}_{n} )] \rangle.
\end{equation}
The $n$-th order participant eccentricity calculated with respect to the
participant plane is defined as:
\begin{equation}
\label{eq7}
\varepsilon_{n} =\frac{\sqrt {\langle r^{2}\cos (n\phi )\rangle ^{2}+\langle r^{2}\sin (n\phi
) \rangle^{2}} }{\langle r^{2}\rangle}
\end{equation}

It has been found that the initial fluctuation causes the difference between the
participant eccentricity and standard eccentricity of the smooth overlap
distribution in previous study \cite{Barrette1994}.

\subsection{Event plane method and resolution}

As known in experiments, the reaction plane angle can not be directly
measured, the anisotropic flows $v_{n}$ are measured with the event plane
method \cite{Poskanzer1998,Ollitrault1998}, which estimates the azimuthal angle of the
reaction plane from the observed event plane determined from the collective
flow itself. The event plane angle is given as$_{\mathbf{:}}$
\[
\psi^{EP}_{n} =\arctan 2(Q_{n,y} ,Q_{n,x} )/n,
\]
where
\begin{equation}
\label{eq8}
Q_{n,x} =\sum\limits_i {\omega_{i} \cos (n\varphi_{i} )} ,
\end{equation}
\[
Q_{n,y} =\sum\limits_i {\omega_{i} \sin (n\varphi_{i} )} .
\]
The above sum goes over all the particles used in the event plane
reconstruction. $\varphi_{i}$ and $\omega_{i}$ are the azimuthal angle
and weight for particle $i$. Since the optimal choice for $\omega_{i}$ is to
approximate $v_{n}$(p$_{T}$,y), the transverse momentum is a common choice
as a weight \cite{Manly2006}. The observed $v_{n}$ measured with respect to the event plane is written by:
\begin{equation}
\label{eq9}
v_{n}^{obs} =\langle \cos [n(\varphi_{i} -\psi^{EP}_{n})]\rangle.
\end{equation}
Since finite multiplicity limits the estimation of the angle of the reaction
plane, the $v_{n}$ has to be corrected by the event plane resolution for
each harmonics given by:
\begin{equation}
\label{eq10}
\Re_{n} =\langle \cos [n(\psi^{EP}_{n} -\psi^{RP} )] \rangle,
\end{equation}
where the angle brackets mean an average over a large event sample. The
event plane resolution depends on the multiplicity of particles used to
define the flow vector and the average  flow of these particles via the
resolution parameter \cite{Ollitrault1998,Poskanzer1998,Danielewicz1985}:
\begin{equation}
\label{eq11}
\Re_{n} (\chi )=\frac{\sqrt \pi}{2}\exp (-\frac{\chi^{2}}{2})(I_{(k-1)/2} (\chi^{2}/2)+I_{(k+1)/2}
(\chi^{2}/2))\quad ,
\end{equation}
where $\chi =v_{n} \sqrt M$ with $M$ the multiplicity and $I_k$ is the modified Bessel function.

To calculate the resolution we divide the full events into two independent
sub-events of equal multiplicity \cite{Ollitrault1993}. The sub-event resolution is defined
as:
\begin{equation}
\label{eq12}
\Re_{n,sub} =\sqrt {\langle \cos [n(\psi_{n}^{A} -\psi_{n}^{B} )]\rangle} ,
\end{equation}
where $A$ and $B$ denote the two subgroups of particles. For the given $\Re
_{n,sub}$ the solution for $\chi$ in Eq.~[\ref{eq11}] is done by iteration. The
full event plane resolution is obtained by:
\begin{equation}
\label{eq13}
\Re_{full} =\Re (\sqrt 2 \chi_{sub} ).
\end{equation}
The final collective flow with respect to the event plane is
\[
v_{n} =\frac{v_{n}^{obs} }{\Re_{n} }.
\]

In event plane method, the event plane is calculated by the final momentum
phase space. The flows extracted through this method may
be affected by the following evolution of the reaction dynamics. In Ref.~\cite{Han2011},
it shows that in Au$+$Au collisions at $\surd $s$_{NN}$  = 200 GeV from a A Multi-Phase Transport (AMPT) the elliptic
flow $v_{2}$ and the triangle flow $v_{3}$ with respect to the event
plane are larger than those with respect to the participant plane. That
illustrates  the evolution of the dynamics does influence the
collective flow.

\section{RESULTS AND DISCUSSION}

With the methods introduced above, we study the collective flows  systematically.
Centrality, transverse component of four velocity and rapidity dependent behaviors
for $v_{n}$ are investigated. Stress are laid on the influences of initial fluctuation,
which is for the first time considered in intermediate energy HIC.
The reaction system Au+Au is investigated to compare with the FOPI experimental data \cite{FOPI-data}.

\subsection{Flows with respect to different planes }

Fig.1 shows y$_{0}$ (left panel) and u$_{t_{0}}$ (right panel) dependencies of
directed flow, elliptic flow and triangular flow
with respect to the {\it reaction plane} ($v_2^{RP}$, i.e. the original Px-Pz plane
in IQMD model itself), {\it participant plane} ($v_2^{PP}$) and {\it event plane}
($v_2^{EP}$) together with the FOPI data  \cite{FOPI-data}, where
y$_{0}$ = $y/y_{p}$ is the scaled rapidity (scaled to the projectile rapidity) and
u$_{t_{0}}$ = $u_{t}/u_{p\,
}$ is the scaled transverse 4-velocity. Note that in $v_{2}$ calculation,
we have applied the detector geometrical cut of the FOPI detector to
make a quantitative comparison (the same for the following parts).

Before the analysis for the influences of initial fluctuation on the flows, we checked the
impact of fluctuation on the particle spectrum itself. It is observed that
there exists the quantitative difference between the average of the sum of $p_x^2$ and $p_y^2$
of particles event-by-event in different initial fluctuation values. However, the effect of initial fluctuation
is almost invisible to single particle spectrum itself. It is understandable that the initial
fluctuation ($x^2$ - $y^2$) in geometrical space have effects on
momentum space ($p_x^2$ - $p_y^2$) event-by-event, but the effect was washed out on
spectrum itself.  It indicates that the effect from initial fluctuation on spectrum are limited at event-by-event level.

Now let's see the directed flow case. It shows that  the y$_{0}$ dependence of
$v_{1}$ with respect to (w.r.t) the participant
plane is weaker  than that w.r.t the reaction plane,
and the shape of $v_{1}$ w.r.t event plane is steeper than that
w.r.t  participant plane. That means the initial fluctuation decreases $v_{1}$, however,
the evolution of dynamics recovers this effect
so that $v_1^{EP}$ is eventually similar to $v_1^{RP}$. From the quantitative comparison with the
data  \cite{FOPI-data}, the directed flow by either the reaction plane method or event plane method
 reasonably reproduces the  $v_1$ data vs rapidity. However, the transverse velocity
 dependence of the $v_1$ data cannot be perfectly fitted by either simulation w.r.t
 various planes. In low u$_{t_{0}}$ region, the $v_1$ simulation with the participant
 plane can describe more or less the data, however, the $v_1$ simulation with the
 reaction plane or event plane gives a close fit to the data.

Secondly, we move on the discussion on rapidity and transverse velocity dependencies of elliptic flow. A
V-shape $y_0$ dependence (Fig. 1c) indicates that proton favors in-plane emission ($v_2>0$) in projectile-like and
target-like regions (larger rapidity) while the squeeze-out ($v_2<0$)
emission dominates in the overlapping region  (mid-rapidity). From mid-rapidity to
projectile-like/target-like rapidity, protons are dominantly emitted from out-of-plane
to in-plane, which is consistent with the decreasing of spectator
shadowing.
Transverse velocity dependence (Fig. 1d) shows that with the increasing of
u$_{t_{0}}$ protons tend to be more squeeze-out emission, i.e. protons with higher
transverse velocity  can be easily ejected from the overlapping zone and more conically focused
perpendicular to the beam axis.

\begin{figure}[htbp]
\centering
\includegraphics[width=9.0cm]{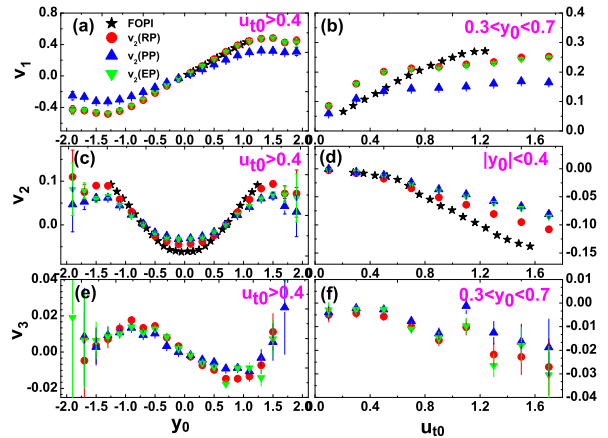}
\caption{\footnotesize (Color online)
Comparison for directed flow $v_{1}$ (a, b), elliptic flow $v_{2}$ (c,d) and  triangular flow $v_3$ (e,f)
with respect to reaction plane (red dots), participant plane (blue up-triangles) and
event plane (green down-triangles) in Au$+$Au collisions
at E = 1$A$ GeV and impact parameter zone 0.25\textless b$_{0}$\textless 0.45 from the
IQMD simulation with the SM parameter. The black stars are the experimental data from FOPI (note the $v_3$ data
of FOPI is not available). The y$_{0}$ dependence integrated over u$_{t_{0}}$ but constrained to
u$_{t_{0}}$ \textgreater 0.4, is plotted on the left panel. The
right panel shows the u$_{t_{0}}$ dependence in the indicated y$_{0\,
}$ region. }
\label{fig1}
\end{figure}

From Fig.1c and 1d, one can see that the $v_{2}$ w.r.t the reaction
plane is a little larger than that w.r.t the participant plane, especially in higher velocity,
which illustrates that the initial fluctuation weakens $v_{2}$. While, there is almost no discrepancy between $v_{2}$ w.r.t
the event plane and that w.r.t the participant plane.
That means the dynamic evolution affects little on $v_{2}$ in the present study. This phenomenon is different from what
is known at RHIC energy, where the initial fluctuation enhances $v_{2}$ and
the dynamic evolution changes $v_{2}$ further \cite{Han2011}. In the figure, the FOPI data
is again plotted in order to check our model calculations. Although these three methods w.r.t
different planes for elliptic flow calculation do not
give full quantitative agreements with the data, the trends of $v_{2}$ as
functions of y$_{0}$ and u$_{t_{0}}$ are the similar with the data. Relatively, $v_2$ of ideal
reaction-plane method approaches to the data nicely.

Thirdly, we also study triangular flow $v_{3}$  with
those three plane methods as functions of y$_{0}$ and u$_{t_{0}}$ as shown in
Figs. 1e and 1f. They show that
the initial fluctuation smooths the shape of the $v_{3}$ dependence on
y$_{0}$ and u$_{t_{0}}$. Just like its effects on $v_{1}$,
the initial fluctuation reduces the amplitude of $v_{3}$. It is also found that the
magnitude of $v_{3}$ w.r.t event plane is not equal to that w.r.t reaction plane.
However, all the magnitudes of $v_{3}$ calculated from various methods are limited only within $\pm$ 2$\%$.
This is very different from the RHIC case, at which triangular flow originates mainly from the initial fluctuation.

\subsection{Impact parameter and beam energy dependence}

\begin{figure}[htbp]
\includegraphics[width=9.0cm]{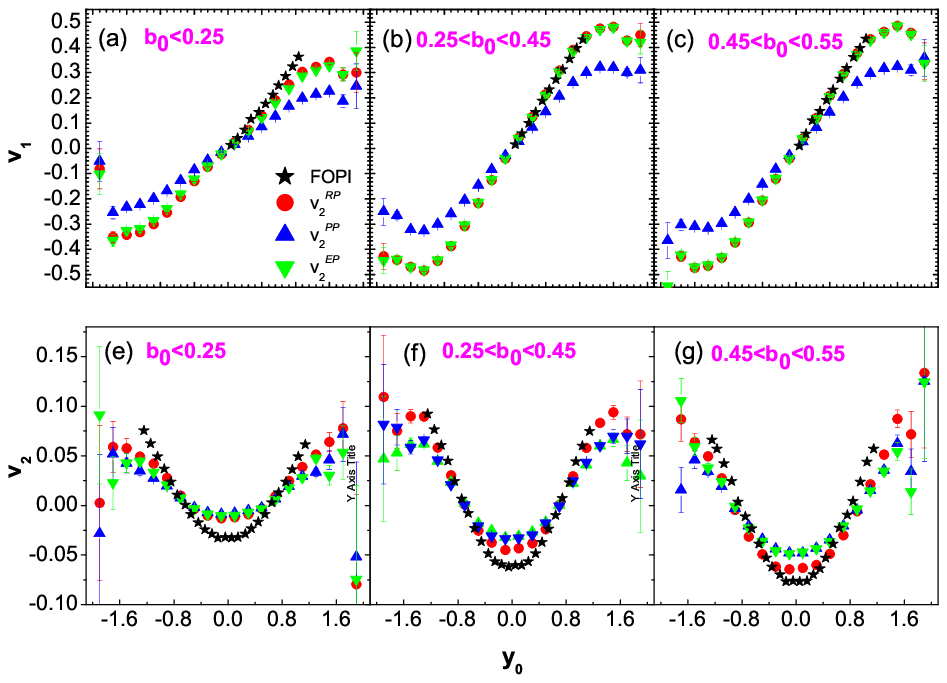}
\caption{\footnotesize (Color online)
Rapidity dependence of directed flow $v_{1}$ (upper panels) and elliptic flow $v_{2}$ (bottom panels) of protons in Au$+$Au collisions for different indicated centrality ranges. Transverse 4-velocities u$_{t_{0}}$
below 0.4 are cut off.}
\label{fig_Nb_dep_v1v2}
\end{figure}

In Fig.~\ref{fig_Nb_dep_v1v2}, we show the centrality dependence of $v_{1}$ and $v_{2}$.
Here the centrality is defined as reduced impact parameter, namely $b_0$,
which is the impact parameter normalized by the largest impact parameter of the system.
Our calculation and FOPI data  \cite{FOPI-data} results have similar values for all centralities.
Both $v_{1}$ and $v_{2}$ reach their maximal at intermediate centrality (0.45$<b_0<0.55$).
The effect of initial fluctuation also shows the largest extent in the same centrality which can be
seen from the increasing difference between $v_1^{RP}$ and $v_1^{PP}$ or between $v_2^{RP}$ and $v_2^{PP}$ .

\begin{figure}[htbp]
\centerline{\includegraphics[width=9.0cm]{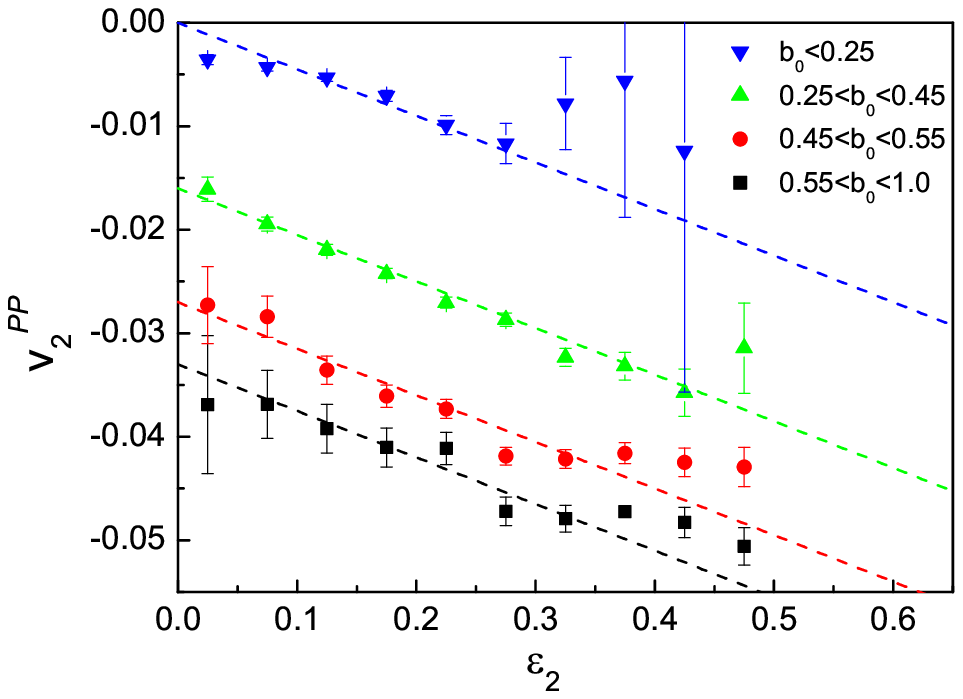}}
\caption{\footnotesize (Color online)
Average elliptic flow with respect to participant plane, $\langle  v_{2}^{PP}\rangle$, as a function of participant plane eccentricity, $\varepsilon_{2}$
at given four impact parameter ranges for Au$+$Au collisions at E = 1$A$ GeV. Line describes a proportional relation of the absolute values of
 $v_2$ versus $\varepsilon_{2}$ with a slope value of -0.045.}
\label{fig_e2}
\end{figure}

The average value of elliptic flow w.r.t participant plane, $v_{2}^{PP}$, as a function of participant $\varepsilon_{2}$,
at mid-rapidity ($|$y$_{0}$$|<$0.5) is presented for 1$A$ GeV Au + Au collisions in
different impact parameter regions in Fig.~\ref{fig_e2}.  It shows that the magnitude of
$v_{2}^{PP}$ is proportional to
$\varepsilon_{2}$. Furthermore, the same slope holds for $v_{2}$  vs $\varepsilon_{2}$ except for a few largest
$\varepsilon_{2}$ points. This indicates that elliptic anisotropy in initial collision geometrical space
leads to an elliptic anisotropy in particle
production along the perpendicular direction in the overlapping participant zone. In previous studies
at relativistic energies, a proportional
relationship between positive $v_2$ and $\varepsilon_{2}$ has also been demonstrated \cite{Han2011}.
Here, a similar proportional trend is, for the
first time, seen for the negative $v_2$ versus $\varepsilon_{2}$. It means that under shadowing
effect, the squeeze-out emission becomes more
 conical when the the geometrical overlapping zone is more prolate. On the other hand, the
 absolute value of $v_{2}$ increases with the impact
 parameter when the value of $\varepsilon_{2}$ is 0, which can be attributed by the increasing
 shadowing effect at larger impact parameter \cite{FOPI-data}.

\begin{figure}[htbp]
\centerline{\includegraphics[width=8.0cm]{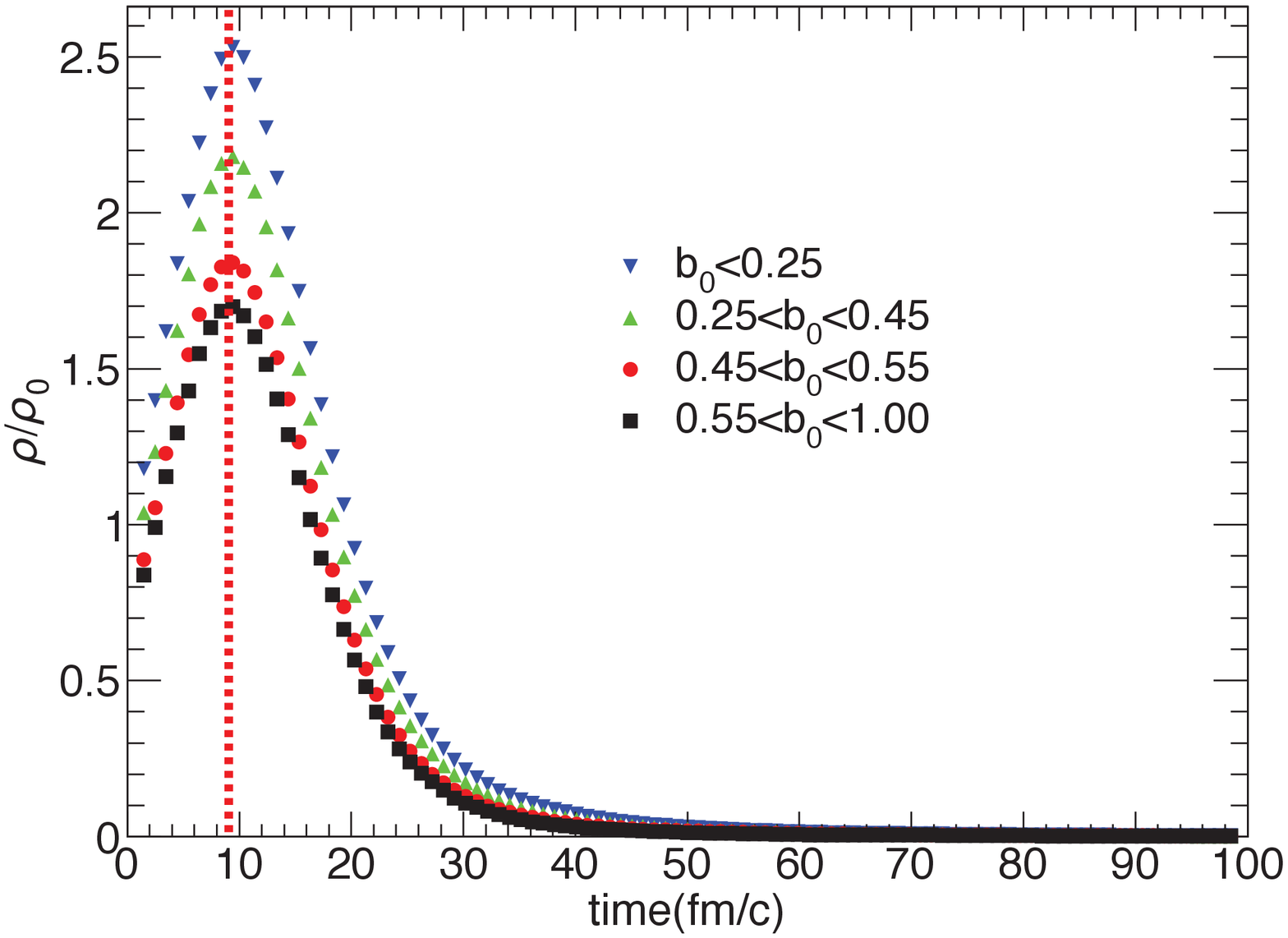}}
\caption{\footnotesize (Color online)
Average density evolutions with time at different impact parameters. A central sphere with radius 5 fm is selected to calculate the density. The maximum compression is around 9 fm/c.}
\label{fig_den_t}
\end{figure}

\begin{figure}[htbp]
\centerline{\includegraphics[width=8.0cm]{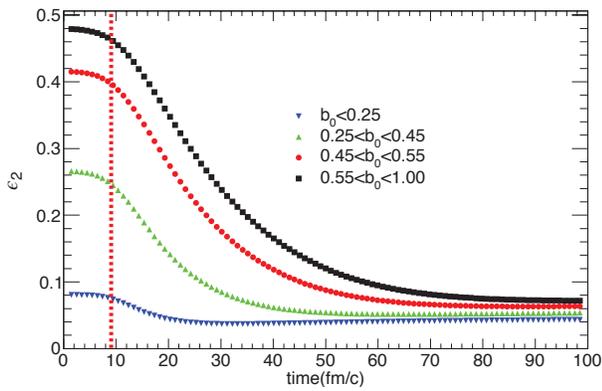}}
\caption{\footnotesize (Color online)
Average $\varepsilon_{2}$ evolutions with time at different impact parameters. The red dash line is around 9 fm/c, }
\label{fig_e2_t}
\end{figure}

In order to understand better the dynamical process,
we consider the time evolutions of the density and $\varepsilon_{2}$. Before the collision (t = 0 fm/c), we set projectile and target at positions
at $-r$ and $r$ respectively, in beam direction in the center of mass system, where $r$ is the radius of Au.
Shown in Fig.~\ref{fig_den_t} are the average density evolutions with time for 1$A$ GeV Au + Au collisions in different impact parameter regions. The system reaches their maximum densities near 9
fm/c, which are almost independent of the impact parameter. Here a central sphere with radius 5 fm
is selected for the density calculation. We then calculate the average eccentricity of the system at this time, because the initial anisotropy changes little before the maximum compression stage. Shown in Fig.~\ref{fig_e2_t} are the time evolutions of the average eccentricity $\varepsilon_{2}$ of the system. It should be noted here
no special rapidity cut is applied here and the eccentricity $\varepsilon_{2}$ is defined according to formula (\ref{eq7}). The eccentricity near 9 fm/c is chosen as the initial eccentricity, and its distribution reflects the initial fluctuation of the system. In this work, we do not focus on
the effect from the distribution of the initial eccentricity, but the effect from the mean eccentricity. The initial fluctuation effect then comes most from the participant plane angle at event-by-event level, which defined in formula (\ref{eq5}).

\begin{figure}[htbp]
\centerline{\includegraphics[width=9.0cm]{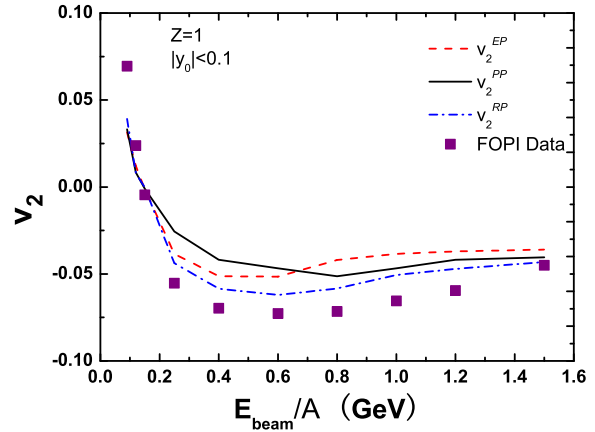}}
\caption{\footnotesize (Color online)
Elliptic flow with respect to different reaction planes as a function of beam energy for $Z$ = 1 particles in
Au +Au collisions at impact parameters of 5.5 fm\textless b\textless 7.5fm. The solid squares represent the
experimental data  \cite{FOPI-data} and different lines for different reaction plane methods.}
\label{fig_Edep_v2}
\end{figure}

Fig.~\ref{fig_Edep_v2} shows a comparison of an excitation function of elliptic
flow  between IQMD calculation results and the FOPI experimental data \cite{FOPI-data}.
Both IQMD calculation and experimental results show a transition energy from positive $v_2$
to negative $v_2$ around 0.2$A$ GeV and a maximum squeeze-out flow around 0.5$A$ GeV and then decreases
towards a transition to in-plane preferential
emission at higher energies which are beyond our present energy points \cite{E895}.
This energy dependence reflects that a competition result from
a consequence of comparable spectator shadowing passing times and fireball expansion times,
in this energy regime \cite{FOPI-data2}.
At energies below 0.2$A$ GeV, particles show positive elliptic flow,
illustrating a collective rotational behavior dominates \cite{Ma-PRC-RC,Ma-PRC2}.
Here, $v_2$ with the participant plane method gives almost same values as the ones with the ideal reaction
plane as well as event-plane, which indicates the initial geometrical fluctuation is actually not significant at lower beam energies.
However, initial fluctuation makes the elliptic flow smaller in absolute value above 0.2$A$ GeV where
a squeeze-out mechanism dominates. As mentioned above, the squeeze-out is an interplay of  fireball expansion
of participant and the shadowing effect of spectator. This  is very different from those cases at
RHIC and LHC energy where stronger positive elliptic flow is mainly developed by the outward pressure in the isolated initial overlapping fireball.
Here the initial fluctuation weakens squeeze-out effect and reduces the elliptic flow, which indicates that
shadowing effect of spectator further quenches the anisotropy induced by the initial geometric fluctuation.
In contrary, the initial fluctuation amplifies the positive elliptic flow at RHIC and LHC energies where the strong outward
pressure further pushes initial fluctuation to become more anisotropic.
In addition, we should mention that the present calculations do not give
a full quantitative fits but underpredict the energy dependence of squeeze-out flow even though
the trend is well described especially for the reaction plane method. Of course, more calculations have been
 performed on flow excitation function  but no consistent agreement between data and calculations exists over
 all the energy range \cite{FOPI-data2}. Hence, some spaces are remained for the model improvement.
Here, in IQMD calculation, again we use the SM equation of state parameter.

\subsection{EOS dependence}

\begin{figure}[htbp]
\centerline{\includegraphics[width=9.0cm]{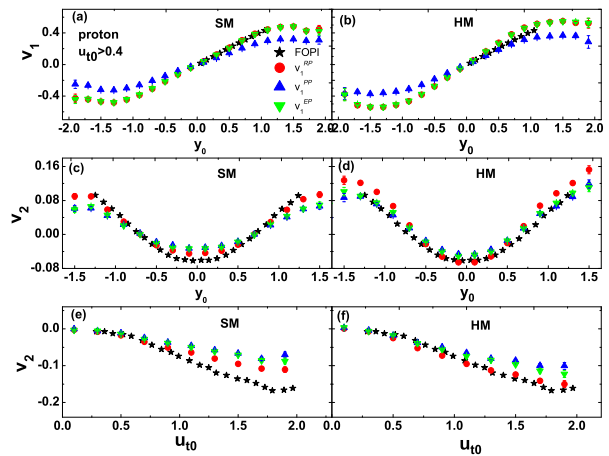}}
\caption{\footnotesize (Color online)
Comparison between directed flow $v_{1}$ and elliptic flow $v_{2}$ from IQMD-SM simulation (left panels) and IQMD-HM simulation (right panels) for Au+ Au
collisions at E = 1$A$ GeV in 0.25\textless b$_{0}$\textless 0.45 centrality
range.  u$_{t_{0}}$ below 0.4 are cut off.}
\label{fig_EOS}
\end{figure}

We also compare the flows with different equation of state (EOS) parameters, i.e. with either the SM or the
HM as shown in Fig.~\ref{fig_EOS}. Our previous study on the particle spectrum has illustrated that
the HM leads to the higher transverse momentum tail than the SM case \cite{LvM}.
In the viewpoint of the effects on flows, we can see the flows with the SM are smaller than the ones
with the HM, which is related a higher transverse momentum tail for the HM \cite{LvM}.
And the effects of the initial fluctuation on $v_{1}$ and
$v_{2}$ from the SM are smaller than that from the HM.
We note that for quantitative fits to the data, the SM is better for the $v_1$ with the reaction-plane
method, but the HM is better for the $v_2$ data with the same method.
Hence, the model is not fully satisfied to describe the data,
which remains the space to improve the model in some ways.

\begin{figure}[htbp]
\centerline{\includegraphics[width=9.0cm]{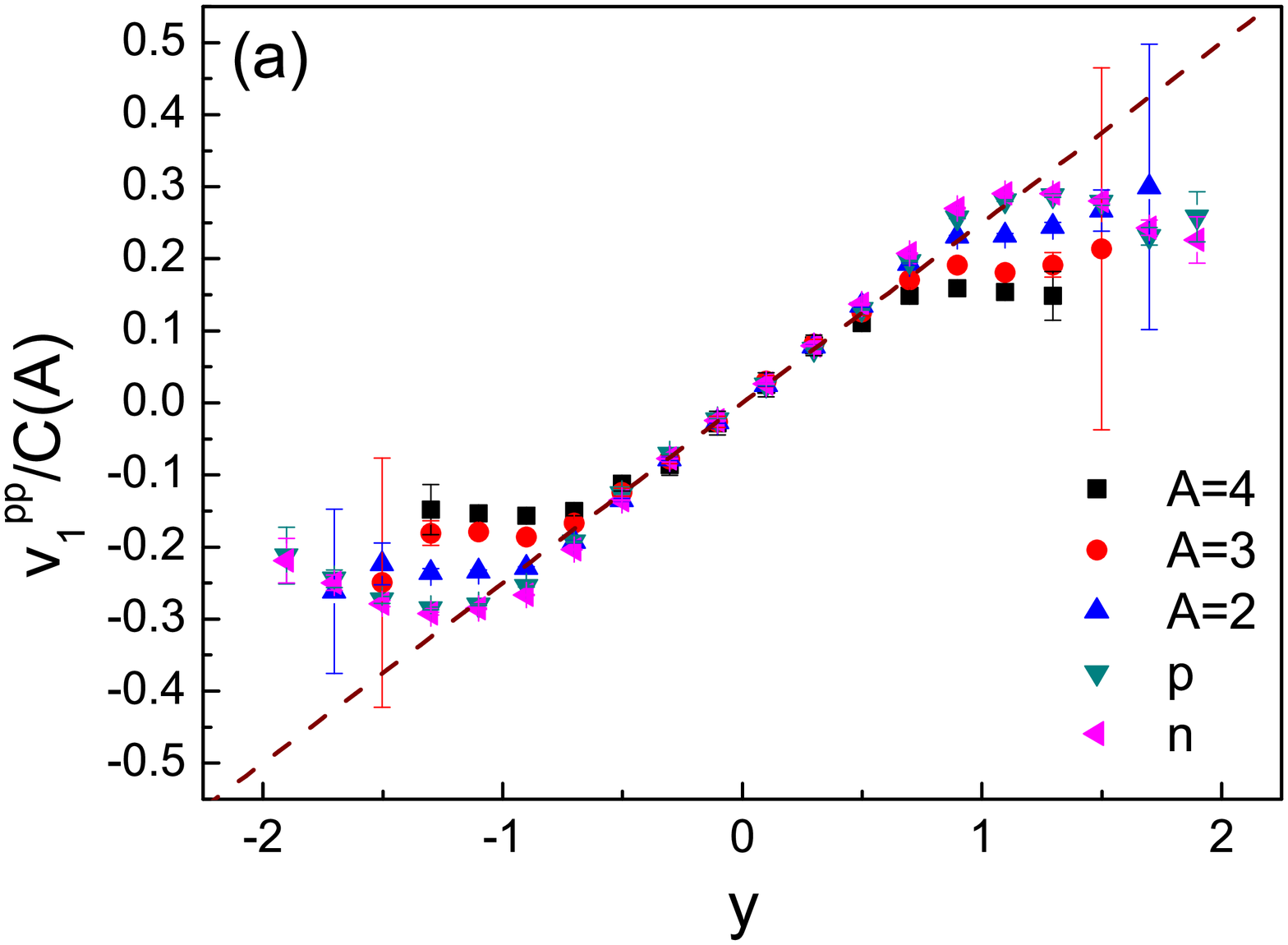}}
\vspace{-0.8cm}
\centerline{\includegraphics[width=9.0cm]{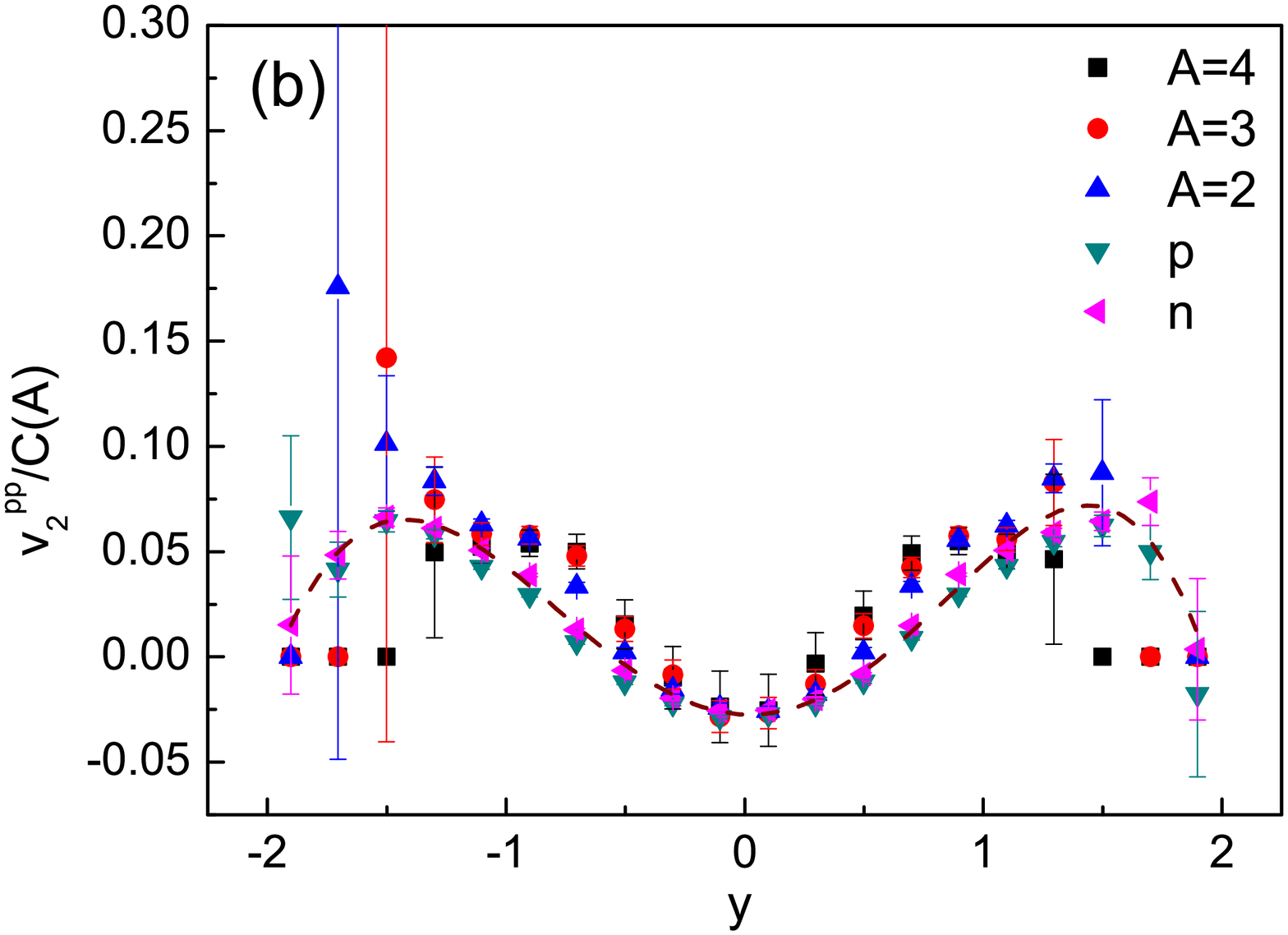}}
\caption{\footnotesize (Color online) Rapidity dependence of the $C(A)$-scaled
$v_1$ (a) and $v_2$ (b) for light particles of Au + Au at 1$A$ GeV, respectively.
Different symbols represent different light fragments. The lines are just for guiding the eyes.
}
\label{v-scaling}
\end{figure}

\subsection{Scaling behavior of flows}

In our previous study on low-energy flow calculation, we have found an approximate mass number scaling holds for directed flow as well as elliptic flow of light nuclear fragments \cite{Yan-PLB,Yan-CP}. Later on, Oh and Ko  also found a similar scaling
of the nucleon number for deuteron at RHIC energies using a  dynamical model \cite{Ko}. Fig.~\ref{v-scaling}(a) and (b) shows the rapidity dependence of $v_1$ and $v_{2}$
scaled by a factor of $C(A)$ for light particles. Here $C(A)=\frac{5}{8}(A+\frac{3}{5}$) is a phenomenological function related to fragment mass number ($A$), where constant term can be seen  originated from the role of random motion and the term with mass number $A$ reflects the collective motion which is proportional to mass. From Fig.~\ref{v-scaling}(a) and (b) we can see for different fragments the $C(A)$-scaled $v_1$ and $v_2$ can merge together especially in mid-rapidity region. It indicates $C(A)$ scaling still works well for $v_1$ and $v_{2}$ after considering initial fluctuation. It should be noted that the participant plane method is adopted
here to consider the effect from initial fluctuation.

\begin{figure}[htbp]
\centerline{\includegraphics[width=9.0cm]{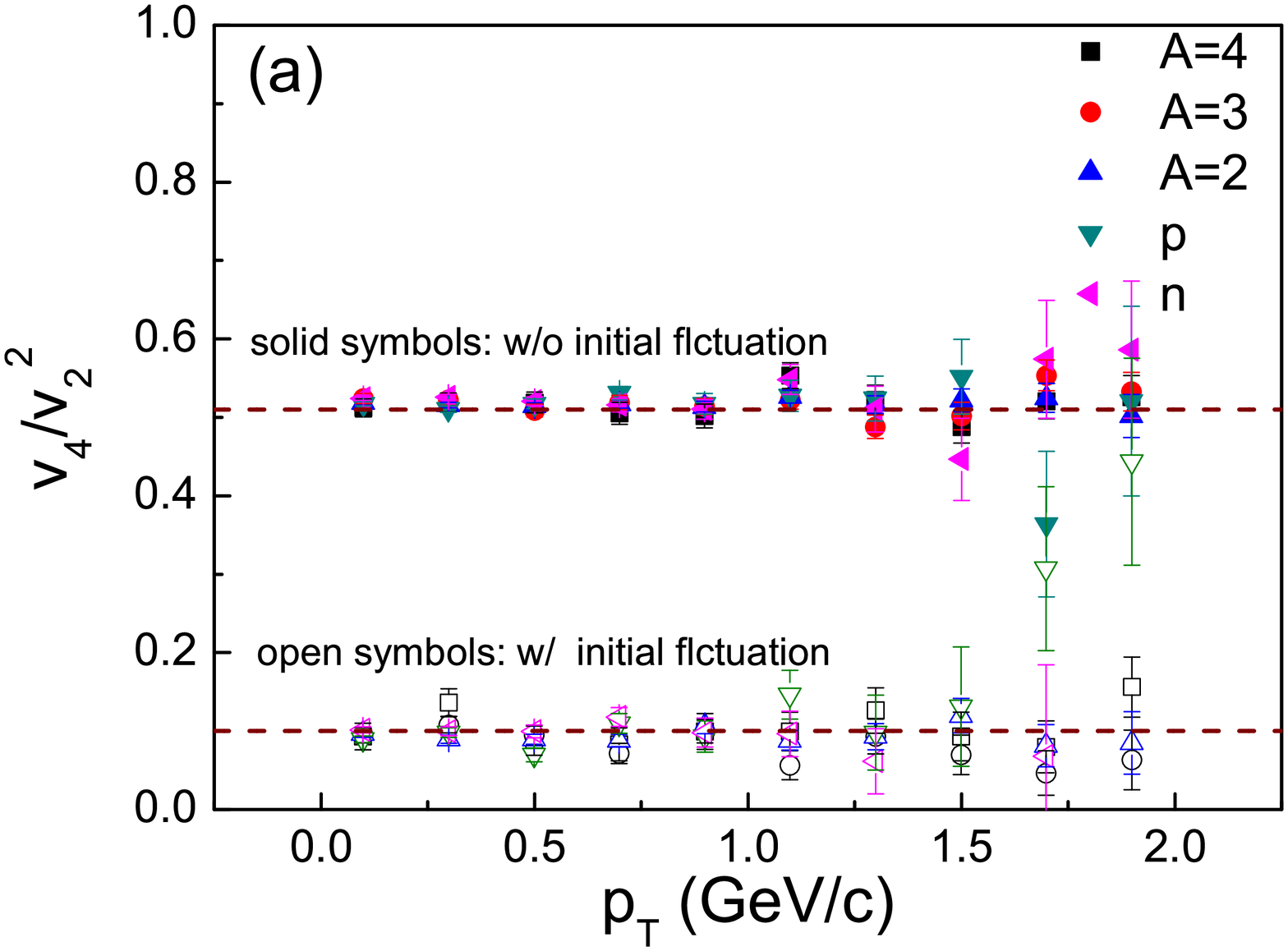}}
\vspace{-0.8cm}
\centerline{\includegraphics[width=9.0cm]{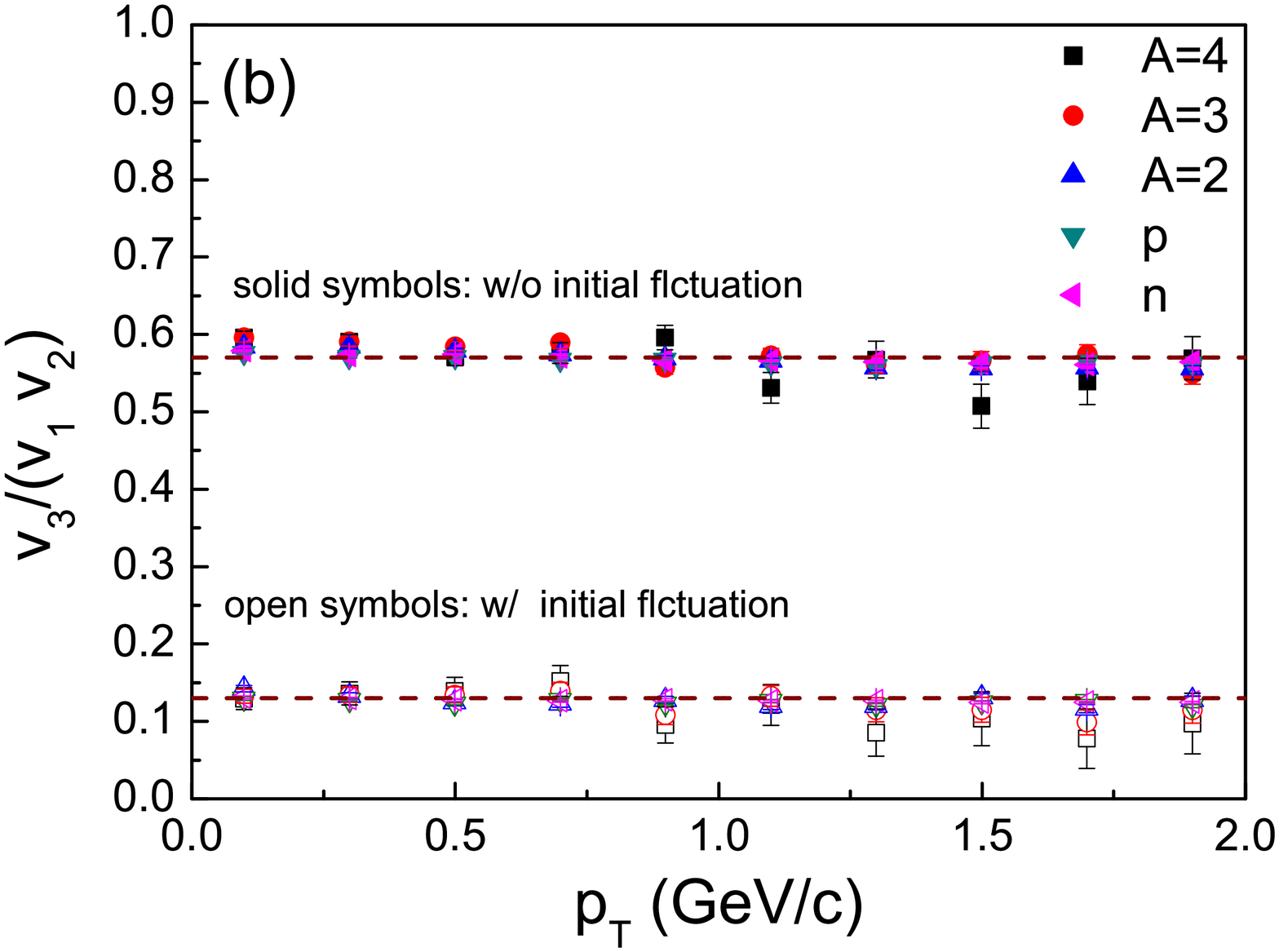}}
\caption{\footnotesize (Color online)
 Ratios of   $v_{4}$/$v^{2}_{2}$ (a) and $v_{3}/(v_{1}v_{2}$) (b)  as a
 function of transverse momentum ($p_T$) for Au + Au at 1$A$ GeV.
 Different symbols represent different fragments as indicted in insert.
 Solid symbols represent for the ratios when the flows are calculated
 without initial fluctuation. Open symbols for the ratios when the
 initial fluctuation is taken into account for flow calculations.
 The lines are just for guiding the eyes.}
\label{fig_v4v22}
\end{figure}

In previous studies, scaling behavior for $v_{4}/v^{2}_{2}$ and $v_{3}/(v_{1}v_{2}$)
has been discussed in hydrodynamical model and partonic transport model at RHIC energies
\cite{kolb,Olli} as well as in Quantum molecular dynamics model in low energy \cite{Yan-CP}.
Here we also address these ratios
as a function of $p_{T}$. Fig. ~\ref{fig_v4v22} (a) and (b) show $v_{4}/v^{2}_{2}$ and
$v_{3}/(v_{1}v_{2}$) of different light particles uptill mass number 4, respectively,
in cases without or with  considering initial fluctuation. Before the initial
fluctuation is taken into account, $v_{4}/v^{2}_{2}$ and $v_{3}/(v_{1}v_{2}$) is about
0.5 and 0.6, respectively, independent of transverse momentum. This $p_T$ independent
behavior indicates a kind of scaling. The value of 0.5 was predicted for $v_{4}/v^{2}_{2}$
by an ideal hydrodynamical model assuming thermal equilibrium \cite{Olli} and the same
ratio was also predicted by the quantum molecular dynamics model in our previous work
on low energy HIC \cite {Yan-PLB} based on the nucleonic coalescence model. Here again,
 $v_{4}/v^{2}_{2}$  is around 1/2 for $A$ = 1-4 particles within IQMD model for 1$A$
 GeV Au + Au collision, which means the nucleonic coalescence model can explain the
 ratio. However, this ratio dramatically decreases after the initial fluctuation is
 taken into account in our flow calculations, eg. the  $v_{4}/v^{2}_{2}$  is only around 0.1.
This behavior seems conversely with the phenomenon in relativistic heavy ion collision
where the high value of  $v_{4}/v^{2}_{2}$  seen experimentally is argued mostly from
by elliptic flow fluctuation \cite{Gombeaud}. However, the ratios for different light
particles still keep the same as well as independent of transverse momentum, i.e. the
scaling behavior dose not break. Similarly,  $v_{3}/(v_{1}v_{2}$) shows a constant
value around 0.6 before the initial fluctuation is considered, but it decreases to
around 0.12 when the initial fluctuation is taken into account. The decreased ratios
indicate that the higher harmonic flow is quenched more strongly due to the initial
fluctuation. This is similar to viscous damping for higher harmonic flows \cite{Han2011}.

\section{Summary}

We analysed the flows as functions of rapidity, transverse 4-velocity,
centrality and beam energy for Au$+$Au collisions at 1$A$ GeV, to investigate
the effect of the initial fluctuation on flows. We would  emphasize
that this is for the first time check of the initial fluctuation effect
on the squeeze-out emission.  Quantitative comparison of the flows w.r.t the participant plane is made
with  the experimental method (event plane method), to investigate the effect of evolution of dynamics.
In addition, we compare the flows with the experimental data from the FOPI Collaboration. We found that
the initial participant fluctuation has indeed some effects on the flows. However,
in contrast with HIC
in the ultra-relativistic region, the initial fluctuation weakens the squeeze-out flow,
which indicates the anisotropy due to initial fluctuation was quenched to some extent by the
spectator shadowing effect.  In addition, squeeze-out flow also
shows proportional to initial eccentricity, indicating squeeze-out essentially develops from the
initial overlapping region.
A quantitative comparison with the excitation function of $v_2$ shows that
our simulation data are smaller
than the experimental data even though the trend is very close, which may be caused by two reasons. One is the
variation of physical parameters (like the ground state
densities,interaction ranges) whose precise values are not known; the other
is the complicate input parameters such as EOS and in-medium cross section etc are not well decided.

In addition, the flow scaling behaviors for different light fragments are also checked in the present simulation.
It is found that the $v_1$ and $v_2$ of different fragments can be scaled by a
function of mass number plus a constant term even  the initial fluctuation has been
taken into account. The ratios of $v_{4}/v^{2}_{2}$ and $v_{3}/(v_{1}v_{2}$)
demonstrate a constant value  which is independent of transverse momentum when the initial
fluctuation is not considered. The value of 1/2 for $v_{4}/v^{2}_{2}$ indicates that the nucleonic
coalescence model can explain  the fragment flows. However, when the
initial fluctuation is considered, the ratios still keep constant values but much less than the
former. It indicates that even though the scaling behavior is not broken,
higher harmonic flows will be much quenched by the initial fluctuation, it is a similar effect of viscous damping.

\vspace{0.8cm}
{\it Acknowledgements:} We thank Dr. Lixin Han and Dr. Zhigang Xiao for helpful discussion.
This work was supported in part by the Major State Basic
Research Development Program in China under Contract No.
2014CB845401, the National Natural Science Foundation of
China under Contracts No. 11035009 and No. 11220101005.

\end{document}